%% file: main.tex
\documentclass[aps,prd,reprint,superscriptaddress,nofootinbib]{revtex4-2}
\usepackage[english]{babel}
\usepackage{amsmath}
\usepackage{amssymb}
\usepackage{graphicx}
\usepackage{color}
\usepackage{lineno}
\usepackage{comment}
\usepackage{tabularx}
\usepackage{lipsum}
\usepackage{orcidlink}

\newcommand\mycolor[1]{\textcolor{black}{#1}}
\newcommand\mycolorbis[1]{\textcolor{black}{#1}}
\newcommand\mycolortris[1]{\textcolor{black}{#1}}
\newcommand\mycolorquater[1]{\textcolor{black}{#1}}

\begin{document}
\title{\mycolorbis{Background of the BULLKID detector array operated with moderate shield on surface}}

\input{bullkid_plugin_authors}

\begin{abstract}
We present the operation with moderate radiation shield in a surface laboratory of BULLKID~\mycolor{(BULky and Low-threshold Kinetic Inductance Detector)}, a \mycolorbis{cryogenic detector} for searches of light Dark Matter \mycolorbis{or Coherent Elastic Neutrino–Nucleus Scattering}.
The detector consists of an array of 60 cubic silicon particle absorbers of 0.34~g each, sensed by cryogenic kinetic inductance detectors.
The analysis presented focuses on data from 15 elements of the array, with two central units used to evaluate the background and with their surrounding elements used as veto. 
The low energy spectrum resulting from an exposure of 290~hours to ambient backgrounds, acquired with the use of external and internal radiation shields, is compatible with the simulations at the level of $(6.8\pm0.4\,{\rm stat.}\pm0.1\,{\rm syst.})\times10^4$~counts~/~keV~kg~days from 2~keV down to an energy of 600~eV. The region between \mycolorbis{225}~eV and 600~eV shows a rise in background in disagreement with the simulations, while not sharing some of the key traits of the low energy excess observed in other cryogenic experiments.
The high energy spectrum \mycolorbis{shape} is in overall agreement with the simulations and displays the typical particle-induced X-ray emission of the surrounding lead.
\end{abstract}

\maketitle

\section{INTRODUCTION}
In the current landscape of Dark Matter searches, WIMP-like particles with mass of \mycolor{order GeV/c$^2$
or below} are interesting candidates~\cite{Bertone2005}, however the direct detection of
such particles requires nuclear recoil detectors with energy threshold of hundreds or
tens of eV~\cite{Albakry2025,Angloher2024,EDELWEISS:2016boq}. 
For \mycolor{these reasons cryogenic particle detectors, thanks to their sensitivity to sub‑keV nuclear recoils are a suitable technology for direct Dark Matter detection} as well as for the search for \mycolor{Coherent Elastic Neutrino–Nucleus Scattering} (CE$\nu$NS)~\cite{Strauss:2017cuu,Augier2024,Ackermann2025,Miner25,Belov2025}, a process with similar low threshold requirement.
In these cryogenic experiments the detector consists \mycolorbis{of} a target crystal acting as particle
absorber, converting the energy of the nuclear recoil into phonons that are then sensed by
semiconducting~\cite{EDELWEISS:2016boq} or superconducting~\cite{SuperCDMS:2017mbc,CRESST2019,Strauss:2017cuu} thermometers.
Existing and future cryogenic experiments face the challenge of deploying a large target mass,
necessary to achieve the needed exposure to probe rare events in a reasonable time frame.
\mycolorbis{In the WIMP search} an effective exploration of interactions with a cross-section lower than 10$^{-40}$~cm$^2$
would require an experiment with approximately one kilogram of active mass and zero background.
\mycolorbis{Background mitigation is also a crucial challenge for cryogenic particle detectors. Even when deployed in a shielded environment designed to mitigate the effects of environmental 
radioactivity, an unexpected excess of events populates the energy spectrum close to
threshold~\cite{excess2022,Angloher2023,Baxter2025}.}

The BULLKID-DM collaboration aims to provide such an experiment by introducing a new detector
\mycolorbis{layout}~\cite{BULLKID-DM}: the device will be composed of \mycolorbis{2320}~silicon dice sensed by cryogenic 
phonon-mediated kinetic inductance detectors (KIDs), for a total active mass of 800~g.
With respect to other solid-state experiments in the
field \mycolorbis{BULLKID-DM} aims to control known and unknown sources of background by creating a fully active array of detectors, with no
inert materials in its structure, as well as by performing a fiducialization of the target.

The detector that BULLKID-DM intends to use in the experiment is based on the technology
developed by the project BULLKID between 2019 and 2023~\cite{bullkid2022}. The BULLKID prototype consists in an
array of 60 silicon dice of 0.34~g in mass, each die sensed by a multiplexed KID, for a total instrumented mass of $60\times 0.34$~g $\simeq 20$~g.
This prototype was operated on surface in the cryogenic laboratory of Sapienza University
of Rome, demonstrating a low energy background flat down to an energy threshold
of 160~eV~\cite{bullkid2024} to a level of \mycolor{$(2.0\pm0.1\,{\rm stat.}\pm0.2\,{\rm syst.})\times10^6$~counts~/~keV~kg~days (DRU).}

On one hand this result \mycolorbis{demonstrated} the self-vetoing capabilities of the BULLKID \mycolorbis{array}, without the
observation of the unexplained low energy excess that is limiting other cryogenic 
experiments~\cite{excess2022,Angloher2023,Baxter2025}.
On the other hand, currently operating experiments make use of extensive \mycolorbis{ambient} background reduction
techniques: to compare the performance of BULLKID with the present experimental landscape
it is necessary to lower the background level. 
\mycolortris{In fact the low energy excess can arise at any level of background: it has been observed by experiments operating at a background level lower than $10^3$~DRU~\cite{Miner25,excess2022,Angloher2023,Baxter2025}, but also by experiments operating at a level of $\sim 10^4$~DRU~\cite{nucleus2026} or greater~\cite{SuperCDMSCPD}. For this reason any reduction in background level achievable with the current experimental conditions represents a valuable improvement to investigate the low energy excess.}
\mycolortris{In this work} we present the operation of the BULLKID detector under moderate shielding
conditions, as well as the resulting background level achieved \mycolortris{by combining the data of multiple sensor clusters and }after applying an
event discrimination analysis, \mycolortris{described in Appendix~\ref{sec:appendix1}, that achieves a higher efficiency} with respect to the procedure already demonstrated in 
Ref.~\cite{bullkid2024}. 

\begin{figure*}[t]
\centering
\includegraphics[width=1.0\linewidth]{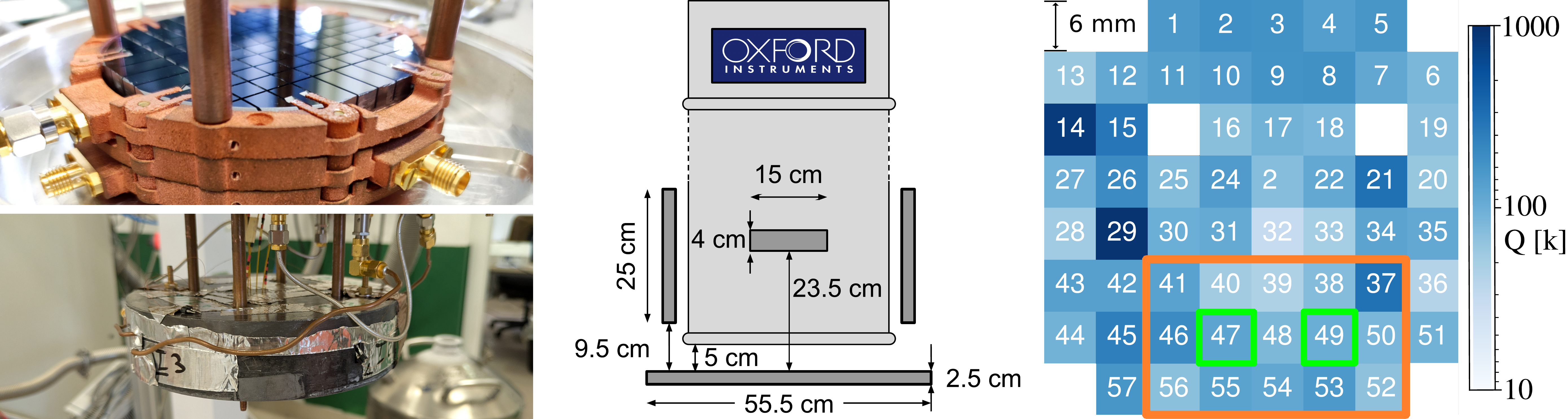}
\caption{
{\bf Top left)} The BULLKID detector operated in this measurement. It consists of a silicon wafer, 3 inches in diameter and 5 mm thick, carved into 60 dice and installed in a copper holder. The two remaining copper holders do not contain any detector and are used solely for mechanical reasons. \mycolorbis{The array of KIDs is on the side opposite to the dice.}{\bf Bottom left)} The detector is placed inside a lead \mycolortris{vessel} for shielding against external radioactivity; {\bf Center)} Schematic representation of the shielding deployed around the detector, in the form of an outer lead castle and an inner lead \mycolortris{vessel}. {\bf Right)} Map of the working KIDs of the detector. The resonators inside the orange continuous line represent the acquired KIDs when the main resonators KID-47 and 49, in the green boxes, trigger. The empty cells represent missing KIDs while the color-scale \mycolorbis{indicates} the total quality factor $Q$ of each resonator.}
\label{fig:detector}
\end{figure*}
\section{EXPERIMENTAL SETUP}
\label{sec:setup}
\begin{figure}[t!]
\centering
\includegraphics[width=1.0\linewidth]{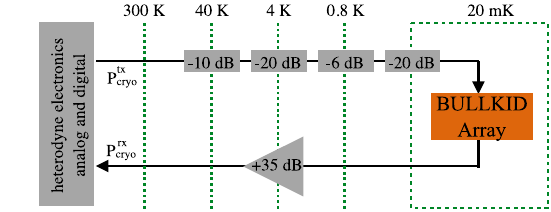}
\caption{\mycolortris{Readout scheme of the detector array placed in the dilution refrigerator.}}
\label{fig:readout}
\end{figure}
The BULLKID detector used for this measurement is shown in Fig.~\ref{fig:detector}~(Top left). The array is installed in a copper holder and is retained by eight forks that are inserted in
corresponding slots carved in the peripheral disk of the wafer. 
The detector and the holder are placed inside a lead \mycolortris{vessel}, shown in Fig.~\ref{fig:detector}~(Bottom left), to provide 4$\pi$ coverage around the array. The cylindrical lead \mycolortris{vessel}, also referred in this work as the ``inner shield'', has a thickness of 4.4~mm on the top and bottom while the circular side wall is 7.0~mm thick. 
\mycolorbis{The inner shield is designed to host up to three
detectors stacked together; in this measurement
however only the top position is occupied by the detector,
with the remaining two copper holders used solely for mechanical reasons.}
An additional Cryophy$^\text{\textregistered}$ \mycolortris{vessel} surrounds the lead case to shield against thermal radiation and external magnetic fields. The detector holder, the inner shield and the Cryophy$^\text{\textregistered}$ \mycolortris{vessel} are anchored to the coldest point of a dry $^3$He/$^4$He dilution refrigerator with base temperature of \mycolorbis{around} 20 mK.
\mycolortris{While the refrigerator and the detector are not stabilized in temperature, small fluctuations around base temperature have a negligible impact on the detector stability~\cite{Cardani:2017qr}.}
The total lead mass available for the inner shield is limited to approximately 2~kg, due to the maximum \mycolortris{weight} limit of the mixing chamber stage of the cryostat.

An external shield \mycolorbis{is} also deployed around the cryostat to further reduce the background of the measurement.
This shield consists of a cylindrical castle of lead bricks surrounding the outer vessel of the cryostat with a thickness of 25~mm. The structure is closed on the bottom by a floor, also made out
of lead bricks, while the top of the cylinder is open due to the spatial constraints imposed by the outer vessel of the cryostat; the overall layout of the internal and external
shields is depicted in Fig.~\ref{fig:detector} (Center).
The total mass of the external shield amounts to $\sim 170$~kg, and is limited by the structural tolerance of the pavement of the cryogenics laboratory, located on the first floor \mycolorbis{of the Physics Department}
of Sapienza University.

The readout is performed with an input coaxial line, running through the cryostat from the outside down to the device. The input line is attenuated at different cryogenic stages for a total of -56~dB, to reduce the \mycolor{effective} noise temperature of the system: \mycolortris{the readout scheme is shown in Fig.~\ref{fig:readout}.} The output line from the device to the outside \mycolorbis{of the cryostat} is amplified by means of an HEMT low-noise amplifier thermalized at the 4 K stage of the cryostat. The microwaves to excite the KID resonators are generated and read back at room
temperature by a commercial Ettus X300 board~\cite{x300_specs} operated with an open-source firmware~\cite{minutolo} customized for the needs of the experiment.

\begin{figure}[t]
\centering
\includegraphics[width=1.0\linewidth]{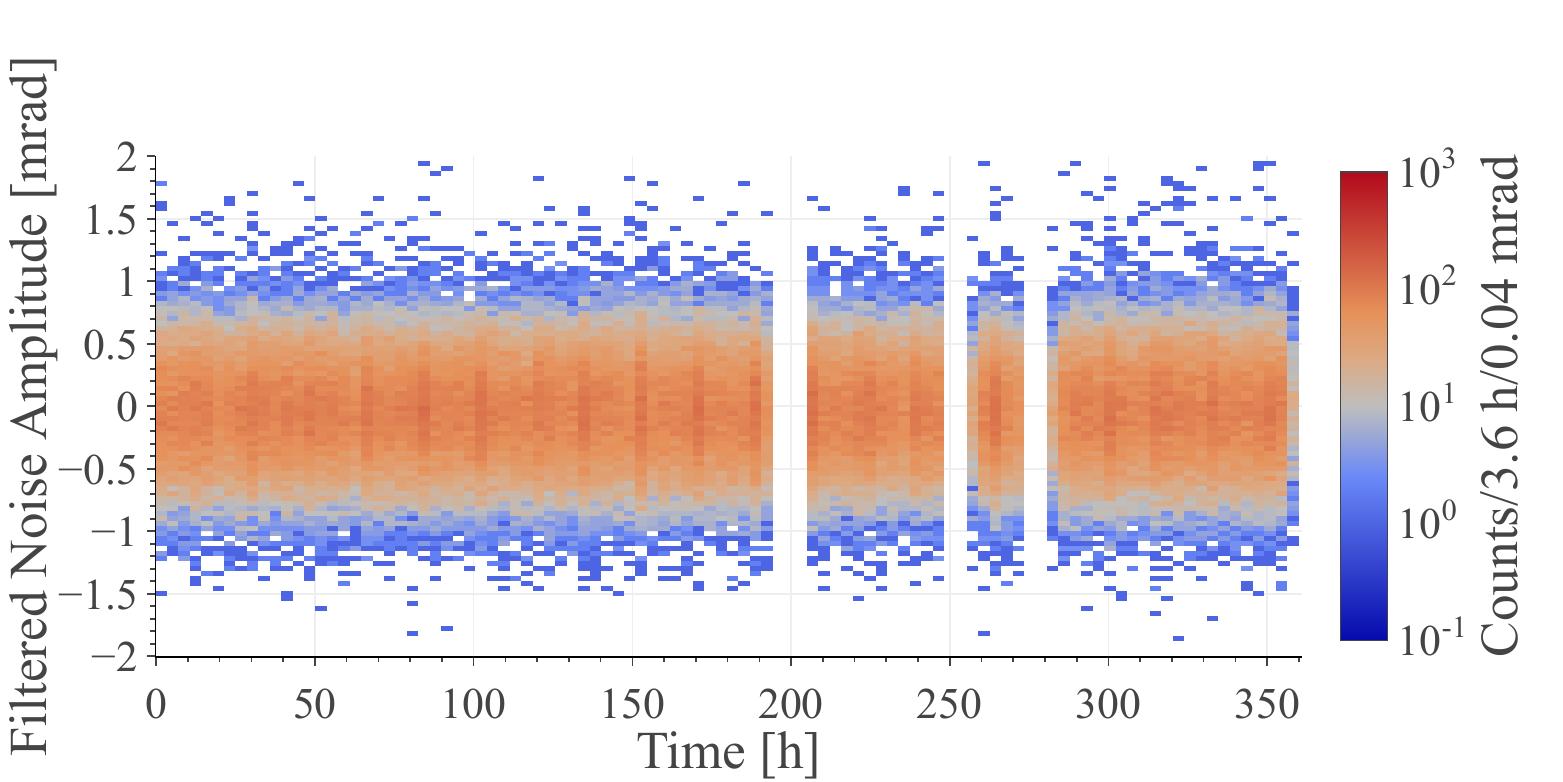}
\includegraphics[width=1.0\linewidth]{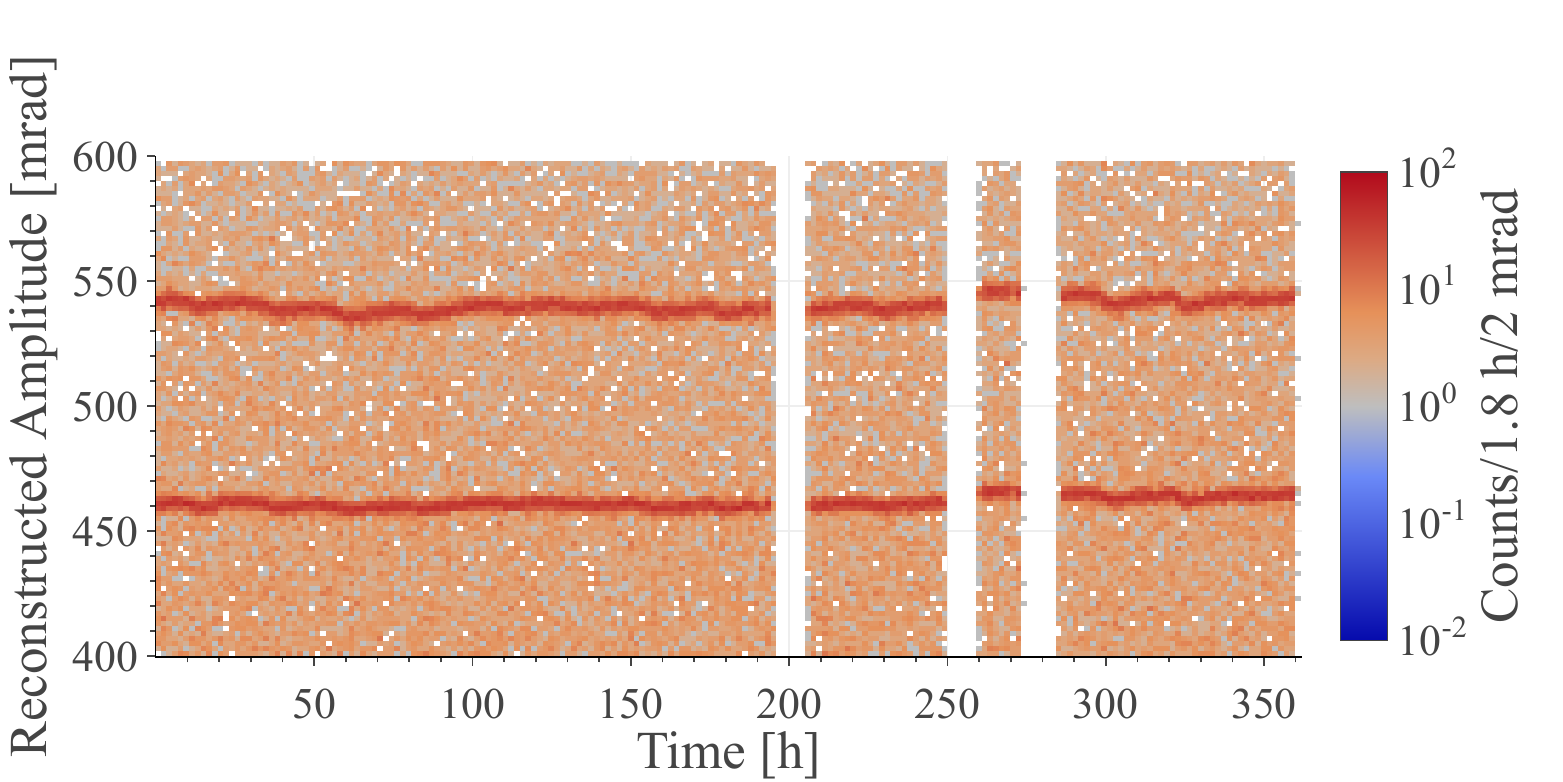}
\caption{{\bf Top)} Distribution of noise samples after applying the matched filter over the whole duration of the data taking for KID-47. The RMS is 0.38~mrad and corresponds to 33~eV after calibration.
{\bf Bottom)} Distribution of KID-47 and 49 pulse response to the phonon leakage produced by energy depositions on the adjacent die of KID-48. Three distinct time intervals in which the signal was not detected have been excluded from the analysis.}
\label{fig:noise&stability}
\end{figure}

During the data-taking, the cluster of 15 KIDs depicted in the orange contour of Fig.~\ref{fig:detector} (Right) is acquired simultaneously when one of the main resonators, KID-47 \mycolor{or} 49, triggers. The goal of the concurrent acquisition is the fiducialization of the active mass, discarding all the events generated by interactions that take place elsewhere from the two main dice~\cite{bullkid2024}. The incoming data stream is filtered online using a matched filter~\cite{Radeka:1966,DiDomizio:2010ph}. Candidate signals are acquired when data exceed a threshold of 4.5 times the standard deviation of the noise. 
The entire data acquisition presented in this work consists in 290 live hours acquired in continuous intervals of 1~h each, separated by a dead time of 3~minutes.
During these dead time intervals samples of noise were acquired by saving a trace every 200~ms, \mycolor{independently of any trigger condition}, to monitor the detector stability over time (Fig.~\ref{fig:noise&stability} Top).

To perform a preliminary energy calibration, optical fibers \mycolorbis{were} excited at room temperature by a 400~nm LED lamp which delivers
controlled photon bursts to the main dice, on the face opposite to the lithography~\cite{Cardani:2021wl,delCastello_2024}. Although the systematic uncertainty associated to the procedure has been evaluated to be lower than 10\%~\cite{Americium}, the detector is cross-calibrated with the \mycolorbis{$\gamma$-rays} and fluorescence rays emitted by the lead \mycolortris{vessel} as reported in Sec. \ref{sec:analysis}. The LED system is also used to provide a signal proxy to monitor the detector response and for the evaluation of the trigger and analysis cuts efficiencies. 
In Fig.~\ref{fig:noise&stability} (Bottom) is shown the reconstructed amplitude by KID-47 and 49 when periodic LED pulses, of constant energy, are fired \mycolorbis{onto} the neighbor die of KID-48 \mycolorbis{exploiting the phonon leakage~\cite{bullkid2024}}. 
The resulting signal is monitored throughout the entire data-taking period to monitor the stability of the detector response for the main dice. Overall we observe both a stable noise and an effectively constant detector response to the test LED signals, with occasional fluctuations $<2\%$. Three time intervals in which the LED did not fire were discarded, as a precaution against undetected fluctuations.
%

\section{BACKGROUND SIMULATIONS}
\label{sec:simulations}
\begin{figure}[tb]
    \centering
    \includegraphics[width=0.485\linewidth]{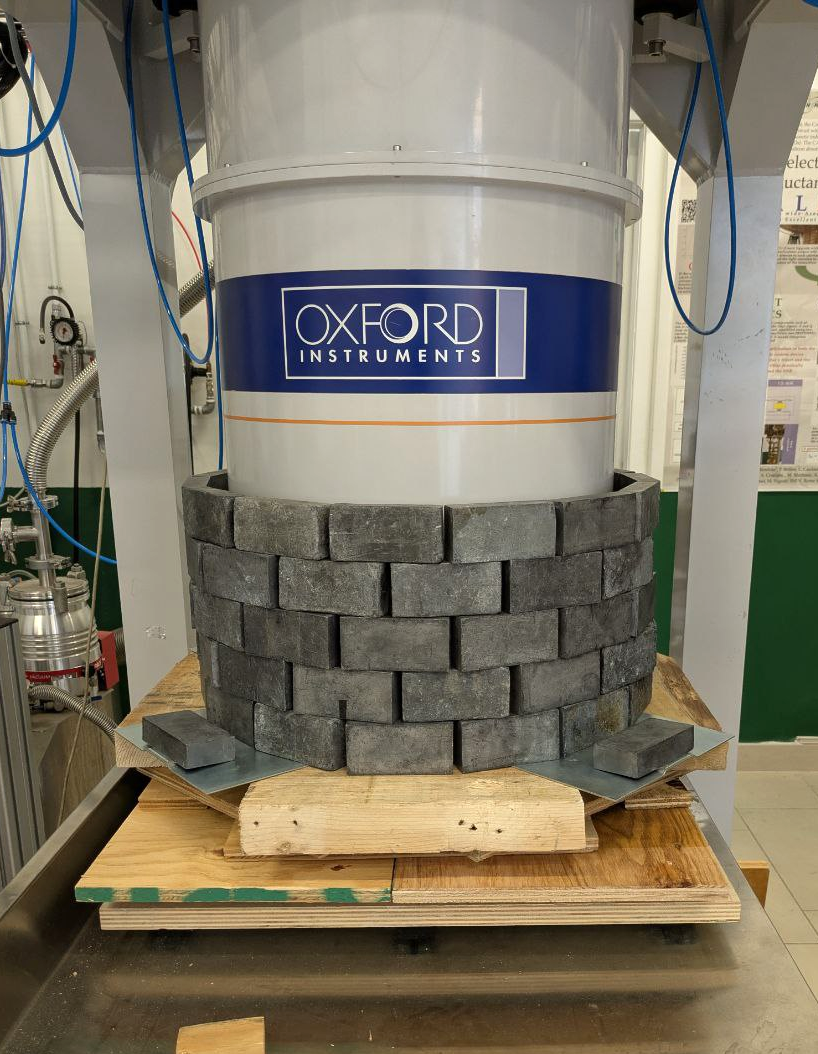}\includegraphics[width=0.515\linewidth]{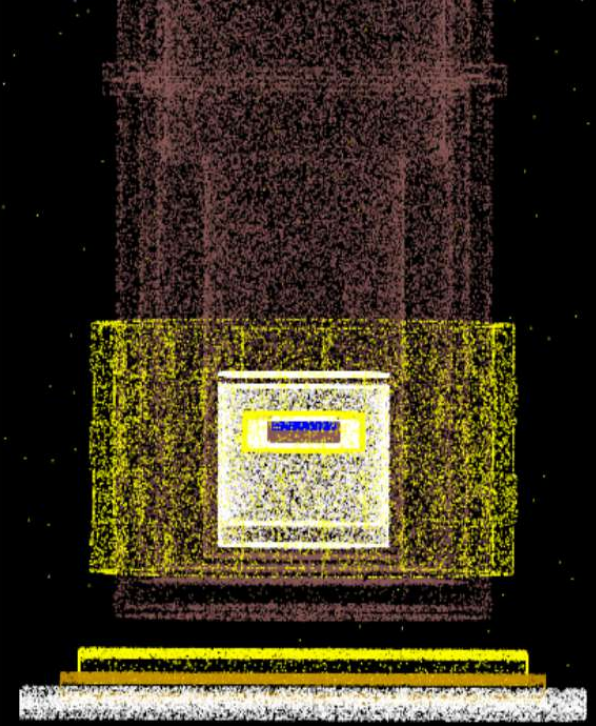}
    \caption{{\bf Left)} View of the external shield surrounding the cryostat in which the detector is installed. {\bf Right)} Geant4 representation of the experiment, where the lead bricks and the support table are shown in yellow. The BULLKID detector is represented in blue.}
    \label{fig:leadbricks}
\end{figure}

The background estimation is performed using the Geant4 toolkit (version 11.1.3) \cite{ALLISON2016186,Allison,AGOSTINELLI2003250} and the \texttt{Shielding} physics list is adopted~\cite{shielding}, as it is recommended for deep-shielding applications and neutron transport. A detailed geometry of the experimental setup is implemented, including all components of the cryostat and their associated support structures. The model also incorporates the lead holder, housing a single detector positioned at the top, as in the experimental configuration described in Sec.~\ref{sec:setup}. In addition, the building of the Physics Department of Sapienza University of Rome is included to account for structural effects on the muon flux.
To reproduce the external shield, lead bricks with dimensions of $2.5\times 10 \times 5~\mathrm{cm}^{3}$ are modeled and arranged around the cryostat in a five-brick stack. The bricks are aligned with the detector center rather than referenced to the floor level. Fig.~\ref{fig:leadbricks} illustrates the corresponding Geant4 representation compared with the external setup. %

\begin{figure}[tb]
    \centering
    \includegraphics[width=1.0\linewidth]{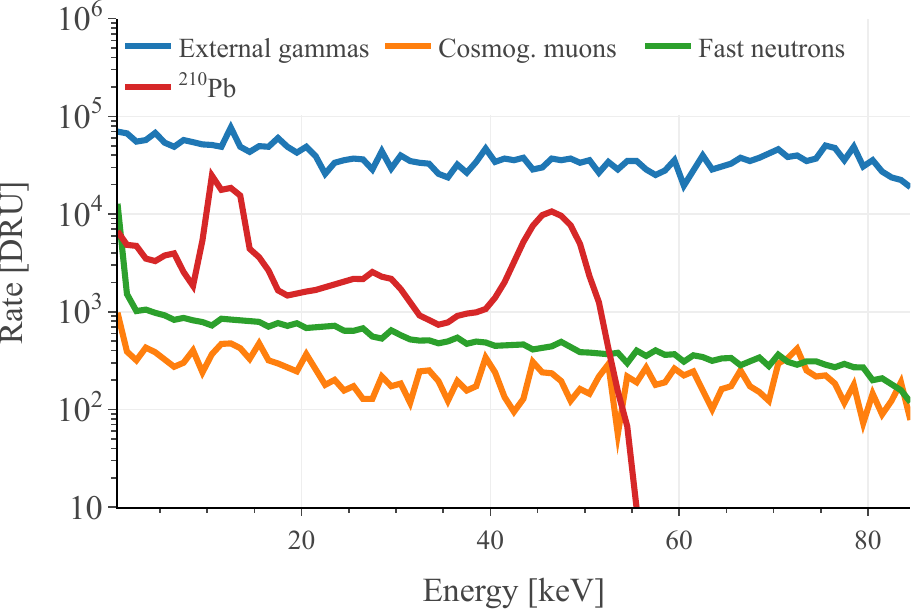}
    \caption{\mycolorbis{Contributions} to the simulated background: the dominant component is represented by the \mycolorbis{$\gamma$-rays} from natural radioactivity, followed by the the contamination of $^{210}$Pb. Sub-dominant contributions are represented by the fast neutrons and cosmogenic muons.}
    \label{fig:simulation}
\end{figure}

The main contributions modeled in this setup are the external \mycolorbis{$\gamma$-rays}, \mycolorbis{the neutron and muon fluxes}, and the contamination from $^{210}$Pb in the inner shield.
The gamma flux is primarily associated with the environmental radioactivity at the site, and was measured using a NaI detector~\cite{ExternalGammas}. 
\mycolortris{The environmental gamma flux was obtained through an unfolding procedure based on a Geant4 simulation of the detector response. Possible biases arising from internal radioactivity of the NaI crystal were investigated. In particular, the contribution from $^{40}$K contamination was determined by placing the crystal inside a thick lead shielding and found to be less than 1\% of the total counting rate. Therefore, internal contamination is not expected to significantly affect the reconstructed gamma flux.}
After spectral unfolding~\cite{UnfoldingBook}, a flux of $(13.47 \pm 2.02)~\mathrm{cm^{-2}\,s^{-1}}$ was obtained in the energy range from $50$ to $3000$~keV. The neutron flux was measured with a DIAMON neutron spectrometer~\cite{POLA2020164078} over a 2.1-day acquisition yielding $(0.010 \pm 0.001) ~\mathrm{cm^{-2}\,s^{-1}}$, with partial contributions of $31\%$ thermal, $35\%$ epithermal, and $34\%$ fast neutrons.~\cite{ByrneJNeutrons}.
For the model discussed in the present work, we only considered the contribution from fast neutrons, as they are the ones with higher energies, and thus are expected to have a major impact on the signal with respect to the thermal and epithermal neutrons. Moreover, we find that neutrons are a sub-leading contributor to the background by approximately two orders of magnitude, as seen in Fig.~\ref{fig:simulation}, which implies that including the thermal and epithermal contribution will not have a significant impact on the overall background estimation.
The muon flux outside the building was simulated using the Cosmic-ray Shower Library (CRY)~\cite{4437209}. The Geant4 model takes as input the flux obtained from CRY to simulate the propagation through the building and the production of secondaries that reach the shield and the detector itself. During the development of the model, no pre-existing measurements of the inner shield radiopurity  were available; therefore, the $^{210}$Pb activity was inferred by comparing the simulated \mycolortris{46.5}~keV peak~\cite{RODRIGUES2016500} with the measured spectrum (see Fig.~\ref{fig:simulation} and Fig.~\ref{fig:highenergyspectrum}). 
A notable feature in the simulated energy spectrum of the current setup is the presence of the particle-induced fluorescence peaks from interactions in the inner shield.
For such low-energy atomic de-excitation processes, several Geant4 PIXE models were tested: Empirical~\cite{PAUL1989105,PAUL199375,ORLIC1994159}, ECPSSR FormFactor~\cite{FFkl,FFm}, and ANSTO~\cite{BAKR202111}. As no significant difference was observed with respect to the default fluorescence model included in \texttt{Shielding}, the latter was chosen due to its lighter computational load.

The output of the simulation is further processed to enable a useful comparison with the data. All the energy depositions happening in the same die and within a time coincidence window compatible with the integration time of the KIDs are summed together. Then, an anti-coincidence algorithm within neighboring dice is applied to mimic the vetoing procedure that is applied to data~\cite{bullkid2024}. Finally, the simulated energy spectrum is convolved with a Gaussian resolution function to reproduce the detector response. The standard deviation of the Gaussian smearing is set to 5\% of the event energy, consistent with the lead-peaks resolution reported in Appendix~\ref{sec:appendix4}, Tab.~\ref{tab:lead_lines}.

Fig.~\ref{fig:simulation} presents the individual contributions of the simulated background. From the comparison of the partial spectra it is possible to observe a subdominant contribution from fast neutrons and muons, contributing to the total spectrum only at the level of a few \mycolor{percentage} points. The main background contribution is therefore represented by the \mycolorbis{$\gamma$-rays} from natural radioactivity and contamination of $^{210}$Pb.

\section{DATA ANALYSIS AND RESULTS}
\label{sec:analysis}
\begin{figure*}[t]
\centering
\includegraphics[width=\linewidth]{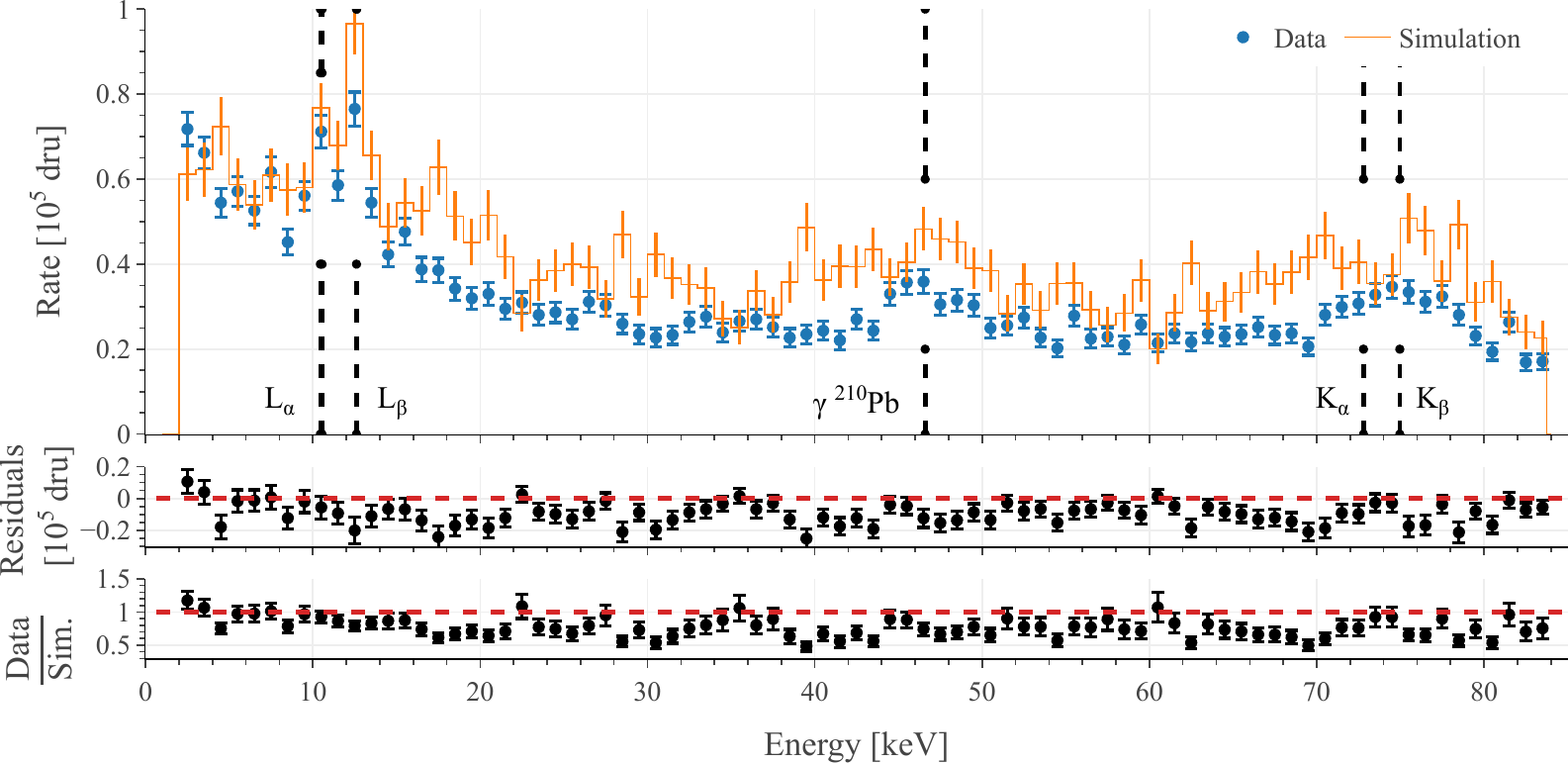}
\caption{ Combined energy spectrum of the two \mycolorbis{main} dice for a total exposure of 290~h $\times$ 0.68~g, from 2~keV to 85~keV
(blue data points). The lead-induced X and \mycolorbis{$\gamma$-rays} listed in Tab.~\ref{tab:lead_lines} are indicated with vertical dashed lines.
The simulated high energy spectrum is superimposed on the data and the bottom panel shows the difference between the measured and the simulated spectra. An energy dependent discrepancy is observed up to a factor of 1.3. 
}
\label{fig:highenergyspectrum}
\end{figure*}
Initial characterization of the array, following the same procedure of Refs~\cite{bullkid2022,bullkid2024}, found a fabrication yield of $\sim 95\%$ with only 3 out of the 60 KIDs being defective, as illustrated in Fig.~\ref{fig:detector} (Right). 

Once particles interact with the silicon target the energy of the nuclear (or electronic) recoil is partially converted into athermal phonons. A fraction of the produced phonons reaches the KID where the deposited energy breaks Cooper pairs into quasi-particles. This process changes the kinetic inductance of the resonator and induces a phase shift in the readout microwave transmitted past the device proportional to the \mycolorbis{deposited energy~\cite{MazinPhD}}.  

Due to the monolithic structure of the BULLKID detector, only $\sim40\%$ of the phonons produced by a particle interaction in a given die are eventually collected by the KID deployed on its surface~\cite{bullkid2022}~\footnote{
In Ref.~\cite{bullkid2022} we compute a total phonon to pair-breaking conversion efficiency $\eta = (24\pm4)\%$: from this result it is possible to estimate the phonon confinement achieved by the geometry of BULLKID. Given the energy to pair-breaking signal conversion efficiency $\eta_0 = 57\%$~\cite{Kozorezov2007}, the corresponding phonon collection efficiency is $\eta / \eta_0  = (44\pm7)\%$.}. \mycolorbis{The rest} of the phonons instead propagate into neighboring units via the common disk, producing a leakage signal. While in principle this effect is unwanted, as it degrades the phonon focusing of each die, it is possible to exploit it to determine if an event originates in a given die or elsewhere~\cite{bullkid2024}.

After applying analysis cuts based on pulse shape variables and the anti-coincidence algorithm, as described in Ref.~\cite{bullkid2024}, we reconstruct the combined high energy spectrum, from 2~keV to 85~keV, of the main KIDs (Fig.~\ref{fig:highenergyspectrum}). The vertical lines indicate the energies of the \mycolorbis{$\gamma$-rays} and fluorescence peaks of lead listed in Tab.~\ref{tab:lead_lines}, which are used for the energy calibration and the evaluation of the analysis efficiencies (see Appendices~\ref{sec:appendix3} and \ref{sec:appendix4}, \mycolortris{and Ref.~\cite{lead-calibration}}). Superimposed to the data we present the expected background from the simulations. The contribution from the contamination of $^{210}$Pb is estimated by fitting the corresponding peak in the data with a Gaussian-plus-linear model and inferring the number of events in the peak. From simulations an activity of $(33.3 \pm 8.3)$~Bq/kg is obtained, consistent with the typical values measured for common lead samples~\cite{SNOlow}. 

A discrepancy in the background level, increasing with energy, up to a factor of 1.3, is observed between the Monte Carlo and the measured spectra: currently the primary source of disagreement under investigation is related to \mycolorbis{systematic uncertainty on the geometry of the simulated setup and on the gamma flux used in the model}. 
Exploring small variations in the structure, such as simulating different offsets in the relative position of the external shield and the detector, did not produce an appreciable effect. \mycolor{We currently do not exclude the possibility that the disagreement in the background level originates from a systematic uncertainty propagating from the assumed gamma flux.} The latter hypothesis will be further tested in future background acquisition campaigns in the same above ground laboratory, since we plan \mycolorbis{to further improve the shielding and to operate} multiple detectors at the same time, as a demonstrator run for the BULLKID-DM experiment.

To improve background rejection in the lower energy region of the spectrum below 2~keV, corresponding to the region of interest of the BULLKID-DM experiment, the event selection has been optimized separately with respect to the high energy spectrum that spans up to 85~keV. 
\mycolortris{The pulse shape selection used for the low energy spectrum employs the same variables used for the high energy spectrum, with stricter cuts. While the pulse shape variables are the ones described in Ref.~\cite{bullkid2024}, the coincidence-based event selection algorithm has instead been refined to achieve high rejection of unwanted events near threshold while retaining an efficiency on signal-like events greater than 45\%; \mycolorbis{for details see Appendix~\ref{sec:appendix1}}.} 

As shown in Fig.~\ref{fig4} (Top), the low energy background detected by the two clusters is compatible, after correcting for the event selection efficiency, which remains constant up to 2~keV. 
The agreement between the results achieved by the two sensors \mycolor{demonstrates the robustness of the analysis procedure} discussed in this work. The data from the two clusters can be combined to double the total exposure to 290~h~$\times$~0.68~g; the combined spectrum is shown in Fig.~\ref{fig4}~(Bottom).

During the data taking the readout was set to trigger on both positive and negative fluctuations \mycolor{of}
the data stream, in order to estimate the rate of noise falsely triggered as signal.
These negative triggered events have been processed using the same analysis and event selection cuts used on regular events. We then define the analysis threshold as the lowest energy bin in the combined energy spectrum where the rate of the negative trigger after the event selection is \mycolorbis{not} comparable with the rate of the main trigger. In the combined spectrum, an analysis threshold of \mycolorbis{225}~eV is obtained, corresponding to $\sim 6$ times the average baseline resolution. 
The combined spectrum depicted in Fig.~\ref{fig4}~(Bottom) is approximately flat in the region from 600~eV and 2~keV at the level of $(6.8\pm0.4) \cdot 10^4$~DRU.
\mycolor{Systematic uncertainties from the energy scale and from the trigger and selection efficiencies contribute an additional} $\pm0.1\cdot 10^4$~DRU. 
The measured spectrum is in agreement within less than 2$\sigma$ with the Monte Carlo model prediction of ($8.2\pm 0.3)\cdot10^4$~DRU.

\mycolor{In contrast,} the region between the threshold of \mycolorbis{225}~eV and up to 600~eV is not
in agreement with the simulation, showing a progressive increase of the count rate 
towards the lower energies that is not explained by the rate of
false positives estimated via the negative trigger. At threshold the count rate is
$5\cdot10^5$~DRU, which is a factor $\sim 7$ \mycolor{higher} than the level of the flat region.

\begin{figure}[t]
\centering
\includegraphics[width=\linewidth]{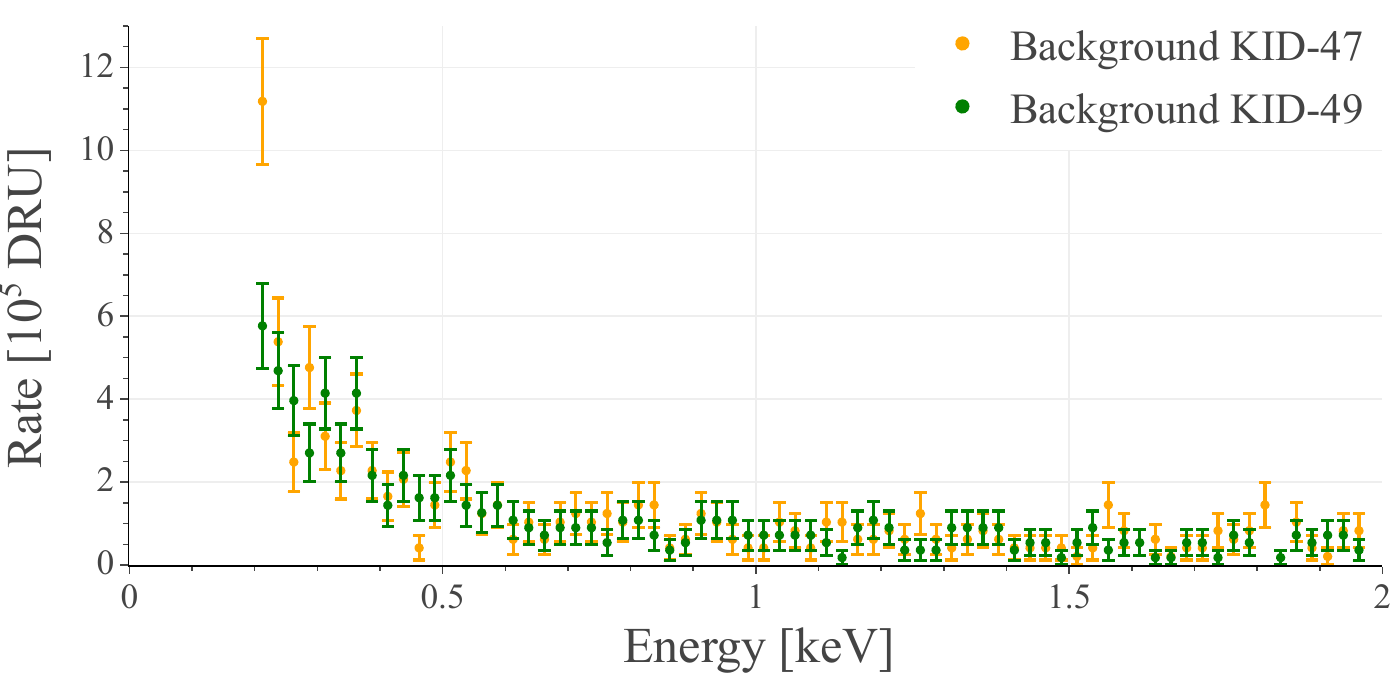}
\includegraphics[width=\linewidth]{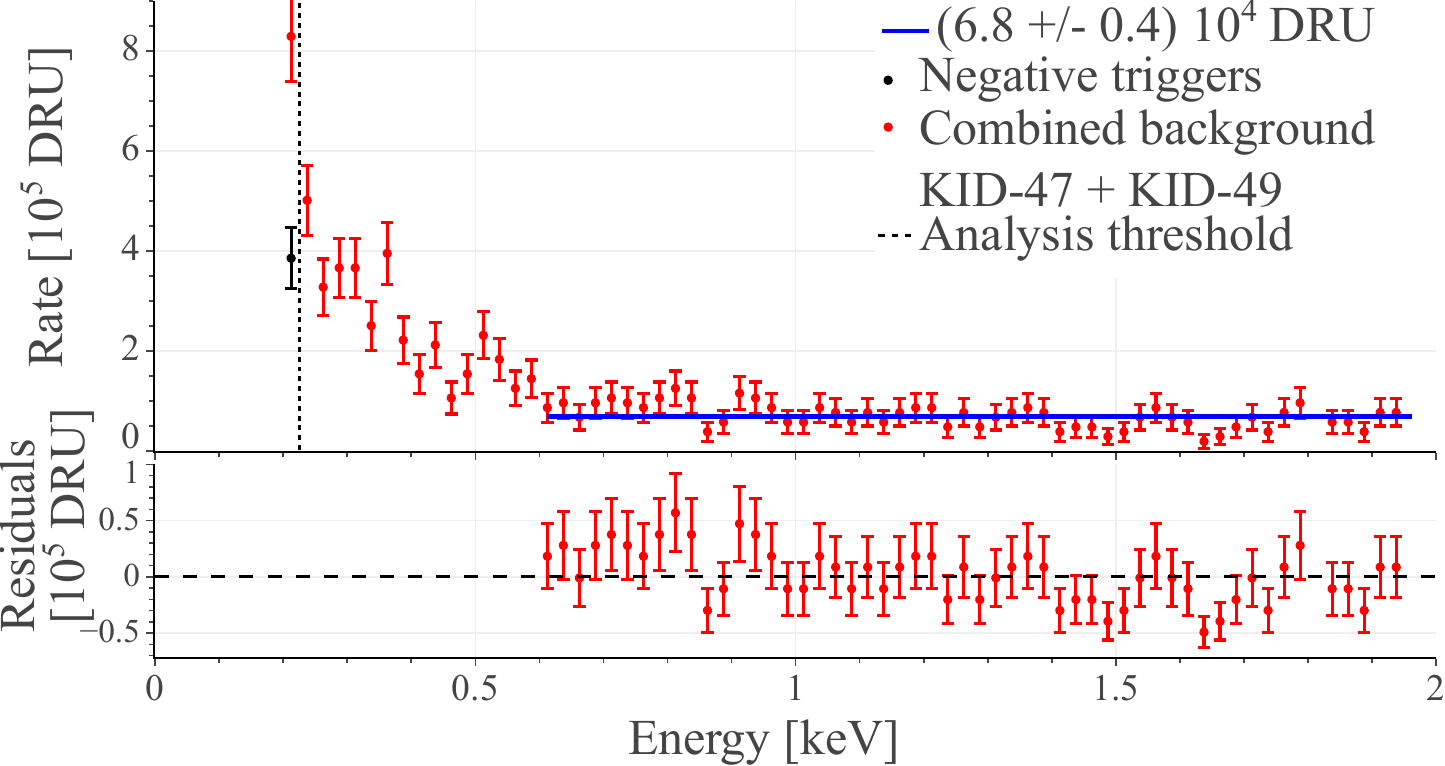}
\caption{ 
{\bf Top:} Energy spectra after 290~h of exposure for the dice sensed by KID-47 (orange) and KID-49 (green). 
After applying the event selection based on the pulse shape discrimination and on the anti-coincidence veto the
two spectra fully overlap, once corrected for selection efficiency.
{\bf Bottom:} Combined energy spectrum of the two \mycolorbis{main} dice for a total exposure of 290~h $\times$ 0.68~g (red data points). 
The energy spectrum is approximately flat down to about 0.6~keV: \mycolortris{fitting the data to a constant yields} a level of $(6.8\pm0.4)\cdot 10^4$~DRU compatible with 
Geant4 simulations. The energy region between 0.6~keV and the analysis threshold of 0.\mycolorbis{225}~keV, displays a progressive increase in the counting rate up to a maximum of $5\cdot10^5$~DRU at threshold. The rate of noise falsely triggered as events is shown by the black point and is not compatible with the observed rise.}
\label{fig4}
\end{figure}

\section{LOW ENERGY RISE}

As previously discussed and as shown in Fig.~\ref{fig4}, the recorded spectrum presents a rise in the counting rate at energies below 600~eV. This rise is currently unexplained by both simulations and noise false positives and thus requires further investigation. At first glance this behavior could resemble the Low Energy Excess (LEE) reported by other experiments~\cite{excess2022,Angloher2023,Baxter2025}; for this reason, the main features of the LEE have been investigated to test this hypothesis. 

\begin{figure}
    \centering
    \includegraphics[width=\linewidth]{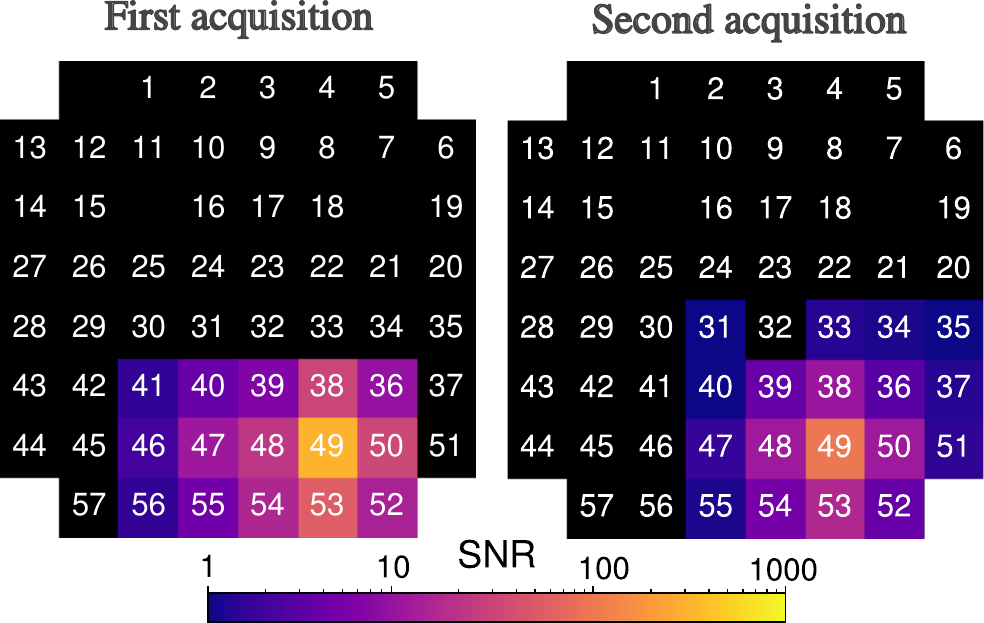}
    \includegraphics[width=\linewidth]{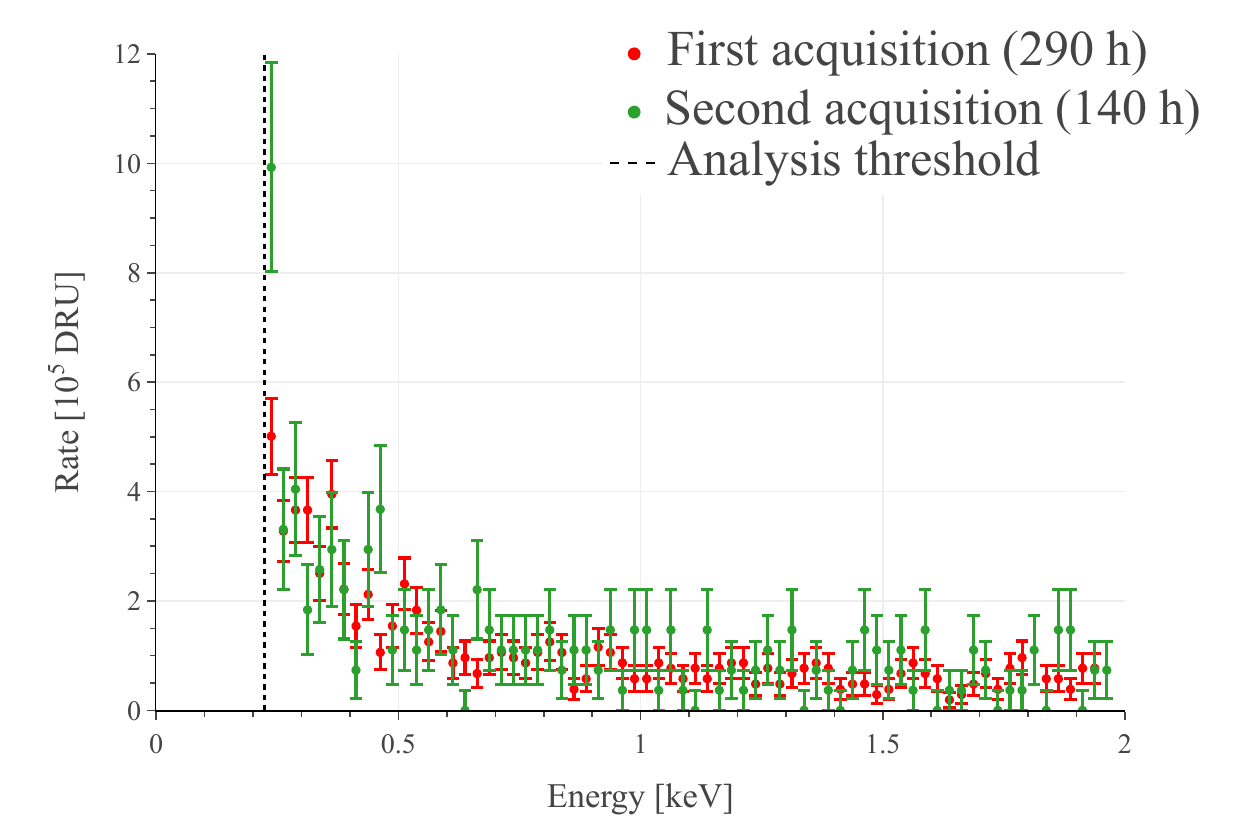}
    \caption{\textbf{Top)} Acquired KID map showing the working conditions of the different KIDs in terms of their SNR (i.e. the optimization of their readout power). The maps are shown for respectively the main (first) acquisition of this work (left), the spectrum is presented in Fig.~\ref{fig4}, and for the repeated (second) acquisition (right) performed to investigate the low energy range. \textbf{Bottom)} Comparison between the spectrum shown in Fig.~\ref{fig4} and the spectrum obtained from the repeated acquisition.}
    \label{fig:second_acquisition}
\end{figure}

A first test consisted in repeating the acquisition under different detector operating conditions: in this measurement the number of the acquired sensors was increased from 15 to 18 KIDs, with only KID-49 triggering the cluster and being used for background evaluation. In this configuration, shown in Fig.~\ref{fig:second_acquisition} (Top), the cluster of sensors used to veto spurious signals is expanded to cover neighbors up to two dice away from the main KID. 
As a \mycolorbis{trade-off}, due to the increased total number of active resonators, it was necessary to decrease the individual bias power of each KID to match the specifications of the readout electronics, resulting in a lower signal to noise ratio (SNR). The SNR is evaluated as the ratio between the amplitude of the background events at the $10.55$~keV lead peak and the noise RMS for each KID. As shown in the bottom panel of Fig.~\ref{fig:second_acquisition} changing the acquisition conditions did not produce any visible change in the recorded spectrum, indicating that the fiducialization technique discussed in Appendix~\ref{sec:appendix1} is robust against reasonable \mycolor{variations in SNR} and configuration of the detector; the observed rise is therefore unlikely to originate from a loss of discrimination power in the analysis. \mycolortris{Since the measurement described in this work we have commissioned an upgraded readout system, able to acquire multiple full arrays with no loss of SNR~\cite{KIT-readout}.}

\begin{figure}
    \centering
     \includegraphics[width=1\linewidth]{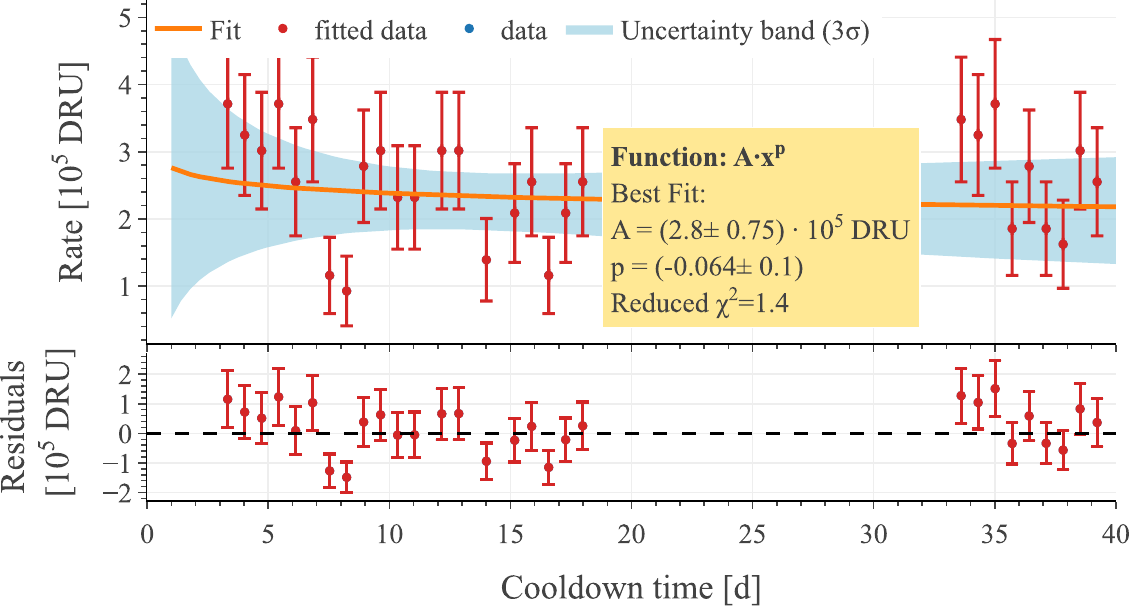}
    \caption{Time evolution of the counting rate measured between the energies of \mycolortris{225~eV and 600~eV.} The time scale starts from the date of cryostat condensation.}
    \label{fig:excess_decay}
\end{figure}

Moreover, many experiments have reported a decaying LEE rate~\cite{Baxter2025,Angloher2023} with time constants ranging from 1 to 100 days. To investigate this time decay, we estimate the time evolution of the counting rate in the energy range from \mycolortris{225~eV to 600~eV}, as shown in Fig.~\ref{fig:excess_decay}.  From the plot no evidence of a decay in the event rate is observed. This conclusion is supported by the result of the \mycolortris{performed power law fit $A\cdot x^p$ which is compatible with $p=0$ within 1.3~$\sigma$.} This behavior is also consistent with the comparison shown in Fig.~\ref{fig:second_acquisition} (bottom), since no difference is observed between the two spectra, acquired starting 3 and 33 days respectively from the time when the system reached base temperature.

From the above considerations and the fact that spectra between the two acquired KIDs are compatible (Fig.~\ref{fig4}, top), the observed rise in the counting rate at low energies appears to be of a different nature with respect to the LEE reported by other experiments, but further investigations in more heavily shielded environments will be carried out by the collaboration.

\begin{figure*}
\includegraphics[width=1\linewidth]{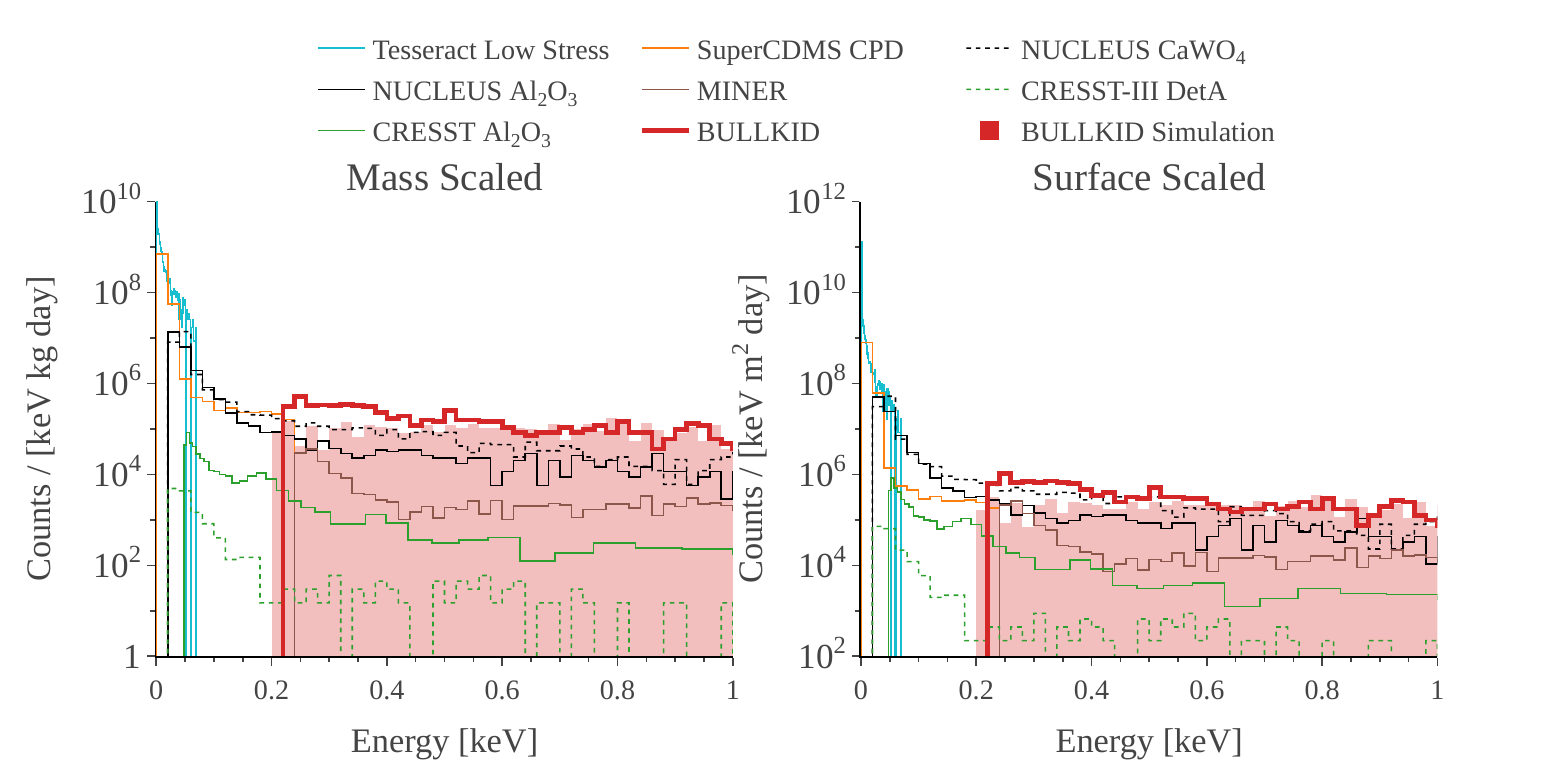}
\caption{Comparison of the BULLKID's spectrum presented in this work (solid red line) with other experiments of the field (see text), alongside the data, we also show the BULLKID Geant4 simulation (shaded red spectrum) to highlight the unexpected low energy rise. The data was partially taken from the public repository maintained by the organizers of the EXCESS workshop and plotted with their tool (see Ref.~\cite{excess2022}, Ref.~\cite{Baxter2025} and \mycolorquater{Ref.~\cite{TesseractLowStress2024}}), while the reactor off spectrum of Ref.~\cite{Miner25} was kindly provided by the Miner collaboration. 
The two panels show the same data \mycolorquater{normalized with respect to the detector mass (left) and with respect to the detector surface (right).}}
\label{fig:comparisonLEE}
\end{figure*}

Finally, it is instructive to compare the spectrum presented in this paper with those recorded by other experiments in the field. \mycolor{Since conflicting observations on the scaling of the excess rate with the target mass have been reported by the CRESST~\cite{Angloher2023} and TESSERACT~\cite{tesseract_mass} collaborations, we present both mass‑scaled and \mycolorquater{surface-scaled} comparisons in Fig.~\ref{fig:comparisonLEE}}. 
\mycolortris{We choose to present the \mycolorquater{surface-scaled} comparison as the mass normalization could hide or distort a potential surface-scaling origin of the LEE.}
As shown from the plots, the spectrum measured in this work exhibits a smaller rise relative to the background level above 1~keV, compared to other experiments \mycolortris{when considering the same energy range of the BULLKID spectrum}. 
\mycolorquater{Moreover, in the mass scaled comparison the measured spectrum is consistent with that reported by SuperCDMS~\cite{SuperCDMSCPD}.}

\section{CONCLUSION and PERSPECTIVES}
In this work we presented the results of the surface background measurement performed with the BULLKID detector operated in a moderately shielded environment. After being surrounded by $\sim 170$~kg of lead, the device was operated for 290~live~hours, acquiring in coincidence two main dice and their neighbors used as veto, for a total of 15 KIDs. The measured background, reduced by a factor 29 with respect to the unshielded configuration~\cite{bullkid2024}, is consistent between the two dice over the full energy range \mycolortris{within $1\sigma$}; the measured spectrum is also in agreement with simulations, with two notable exceptions. 

In the high energy region up to 85~keV the simulation overestimates the background by an energy dependent factor up to 30\%; investigation of this discrepancy is ongoing, with indications pointing towards a systematic uncertainty on the gamma flux.
A second discrepancy between data and simulation is observed in the low energy region, between 600~eV and the analysis threshold of \mycolorbis{225}~eV. While the simulated background in this region is flat, the measured spectrum exhibits an unexplained rise that increases by up to a factor 7 towards threshold. Although this behavior resembles the low energy excess seen by other cryogenic experiments, the absence of the typical decay in counting rate over time and other considerations discussed in this work suggest that its origin is likely different.

Currently, we are investigating a possible particle origin of this rise by extending the simulation model. We also plan above ground measurements with increased shielding, as well as an underground run at Laboratori Nazionali del Gran Sasso targeting a background level of $\sim 10^4$~DRU \mycolorbis{or lower}, to verify whether the discrepancy persists at lower background levels.

\begin{acknowledgments}
\mycolorbis{This work was supported by the INFN, Sapienza University of Rome and co-funded by the European Union (ERC, DANAE, Grant No. 101087663). We further acknowledge the support of the Doctoral School \emph{“Karlsruhe School of Elementary and Astroparticle Physics: Science and Technology”}, the SECIHTI Project No. CBF-2025-I-1589 and DGAPA UNAM Grants No. PAPIIT IN105923 and IN102326.}  Views and opinions expressed are however those of the author(s) only and do not necessarily reflect those of the European Union or the European Research Council. Neither the European Union nor the granting authority can be held responsible for them. We acknowledge the support of the PTA platform for the fabrication of the device. We thank A. Girardi and M. Iannone of the INFN Sezione di Roma for technical support, and the MINER collaboration for providing access to the data.
\end{acknowledgments}
\bibliographystyle{apsrev4-2}
\bibliography{calder}

\begin{appendix}


\section{LEAD-INDUCED ENERGY CALIBRATION}\label{sec:appendix3}
The preliminary LED calibration of the main KIDs is cross-checked using the lead-induced lines listed in Tab. \ref{tab:lead_lines}. These lines are reconstructed by fitting the energy spectrum with a linearly decreasing background component and a Gaussian model for each line in the energy intervals $\left[2,21\right]$, $\left[30,60\right]$, and $\left[60,100\right]$ keV, \mycolortris{following the same procedure of Ref.~\cite{lead-calibration}}. Within each interval, the Gaussian standard deviations are constrained to be identical, while the means and the number of events are fixed relative to one another according to the values reported in~\cite{X_fluo}.
The results from the fit of each Gaussian are summarized in Tab.~\ref{tab:lead_lines}.
\begin{table*}[bt]
\centering
\renewcommand{\arraystretch}{1.4} 
\begin{tabular}{@{}c|c|c||c|c|c||c|c|c@{}}
\hline
 &  &  & \multicolumn{3}{c||}{KID 47} & \multicolumn{3}{c}{KID 49} \\
\hline
Line & $E_{\text{nominal}}$ [keV] & Intensity & n. events & $E_{\text{mean}}$ [keV] & $\sigma/E_{\text{mean}}$ [\%] & n. events & $E_{\text{mean}}$ [keV] & $\sigma/E_{\text{mean}}$ [\%]\\
\hline
$L_{\alpha 2}$ & 10.45 & 11 &$10 \pm 2$ & $9.2\pm0.1$ & $5.3\pm0.9$ &$10\pm2$ & $10.6\pm0.1$  & $4.2\pm0.7$\\
$L_{\alpha 1}$ & 10.55 & 100 &$88\pm15$ & $9.3\pm0.1$ & $5.3\pm0.9$&$95\pm1$ & $10.7\pm0.1$ & $4.2\pm0.7$\\
$L_{\beta 1}$ & 12.61 & 66 &$58\pm10$ & $11.1\pm0.1$ & $4.4\pm0.8$ &$63\pm10$ & $12.8\pm0.1$ & $3.5\pm0.6$\\
$L_{\beta 2}$ & 12.62 & 25 &$22\pm4$& $11.1\pm0.1$ & $4.4\pm0.8$ &$24\pm4$& $12.8\pm0.1$ &$3.5\pm0.6$\\
\hline
$\gamma~^{210}$Pb & \mycolortris{46.54} & 100 &$164\pm38$ & $40.2\pm0.5$ & $5.0\pm1.1$ &$126\pm30$ & $46.3\pm0.4$ & $3.7\pm0.8$\\
\hline
$K_{\alpha 2}$ & 72.80 & 60 &$113\pm12$& $63.5\pm0.3$ & $3.7\pm0.5$ &$129\pm17$& $71.7\pm0.5$ & $4.9\pm0.7$\\
$K_{\alpha 1}$ & 74.97 & 100 &$189\pm21$ & $65.4\pm0.3$ & $3.7\pm0.5$ &$214\pm28$& $73.9\pm0.5$ & $4.7\pm0.7$\\
$K_{\beta 3}$ & 84.45 & 12 &$23\pm3$& $73.7\pm0.4$ & $3.2\pm0.4$ &$26\pm3$& $83.2\pm0.6$ & $4.2\pm0.6$\\
$K_{\beta 1}$ & 84.94 & 23 &$43\pm5$ & $74.1\pm0.4$ & $3.2\pm0.4$ &$49\pm6$ & $83.7\pm0.6$ & $4.2\pm0.6$\\
$K_{\beta 2}$ & 87.32 & 8 &$15\pm2$& $76.2\pm0.4$ & $3.2\pm0.4$ &$17\pm2$ & $86.1\pm0.6$ & $4.2\pm0.6$\\
\hline
\end{tabular}
\caption{Main fluorescence lines of lead and the \mycolorbis{$46.54$} keV $\gamma$-line from $^{210}$Pb contamination. Nominal energy and relative intensity from Ref.~\cite{X_fluo} are listed, along with the number of events, the mean and the relative width for the main KIDs, obtained from the fitting procedure explained in Appendix \ref{sec:appendix3}.}
\label{tab:lead_lines}
\end{table*}
By comparing the inferred mean of the first Gaussian in each energy interval with its nominal energy, we evaluate the response function of the main KIDs with a parabolic model of the form:
\begin{equation}
    E_{\text{mean}} = a \cdot E_{\text{nominal}} \cdot (1+b\cdot a \cdot E_{\text{nominal}} )
    \label{eq:resp_func}
\end{equation}
where the fitted parameters for the two main KIDs are:
\begin{align*}
    a_{47}& =0.90\pm0.01 & b_{47}=-(0.9\pm2.0)\cdot10^{-4}~\text{keV}^{-1} \\
    a_{49}& =1.02\pm0.01 & b_{49}=-(4.9\pm1.4)\cdot10^{-4}~\text{keV}^{-1}
\end{align*}
This procedure validates the standard LED calibration, as described in Ref.~\cite{Americium}, to an accuracy better than 10\%.
Moreover we assess that the residual non-linearity of the LED calibration is at most $5\%$ at $85$ keV.
The calibrated energy $E_{\text{cal}}$ of the main KIDs is obtained by inverting Eq. \ref{eq:resp_func}:
\begin{equation}
    E_{\mathrm{cal}} = \frac{-1 + \sqrt{1 + 4 \cdot b \cdot E_{\mathrm{mean}}}}{2 \cdot a \cdot b}
\end{equation}

\section{EVALUATION OF THE HIGH ENERGY ANALYSIS EFFICIENCY}\label{sec:appendix4}
\begin{figure}
    \centering
    \vspace{0.7 cm}
    \includegraphics[width=\linewidth]{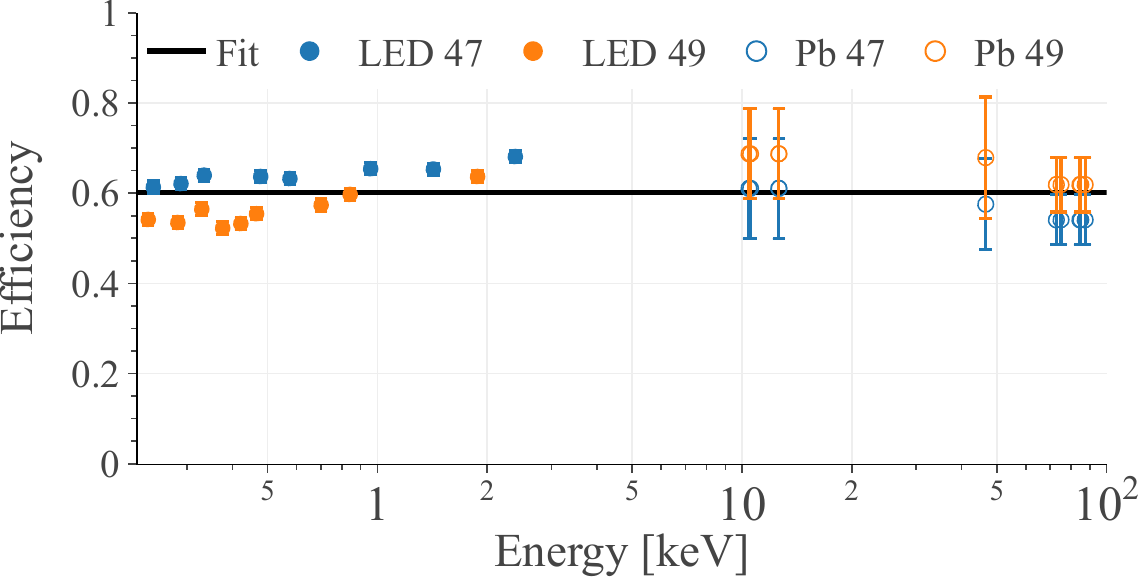}
    \caption{Combined efficiencies (both trigger and analysis) of the main KIDs. The filled symbols are the efficiencies evaluated for low energies (below 2 keV) using the LED pulses. The open symbols are the efficiencies evaluated by comparing the lead-induced lines in the energy spectrum with and without the analysis cuts. \mycolortris{We fit for a flat efficiency over the full energy range, compatible between the two main KIDs, yielding $\epsilon = (60\pm1)~\%$.}}
    \label{fig:efficiecy}
\end{figure}

In Ref.~\cite{Americium}, it was shown that LED pulses \mycolortris{do not represent an appropriate template of known-good} high-energy events in BULLKID due to their reduced penetration depth in the silicon substrate. This results in a different phonon-leakage distribution in neighboring dice that affects the veto analysis. Therefore, unlike in the low-energy region, LED pulses cannot be used to evaluate the efficiencies of the analysis cuts, as done in Sec.~\ref{sec:analysis} and Ref.~\cite{bullkid2024}. The strategy adopted in this work is to evaluate these efficiencies by fitting the lead-induced lines in the energy spectrum. The lines are fitted using the same procedure described in Appendix \ref{sec:appendix3}, both with and without the application of the analysis cuts. By comparing the number of events in both cases, we were able to infer the efficiency of the cuts. Fig. \ref{fig:efficiecy} shows, for both the main KIDs, the efficiencies evaluated via the lead peaks (open symbols) with respect the ones evaluated through LED at low energies (filled symbols). By applying the same analysis cuts we obtain compatible efficiencies of the value of $\epsilon = (60\pm1)~\%$.

\section{CLUSTER EVENT SELECTION}\label{sec:appendix1}
\begin{figure}[tb]
\includegraphics[width=1.0\linewidth]{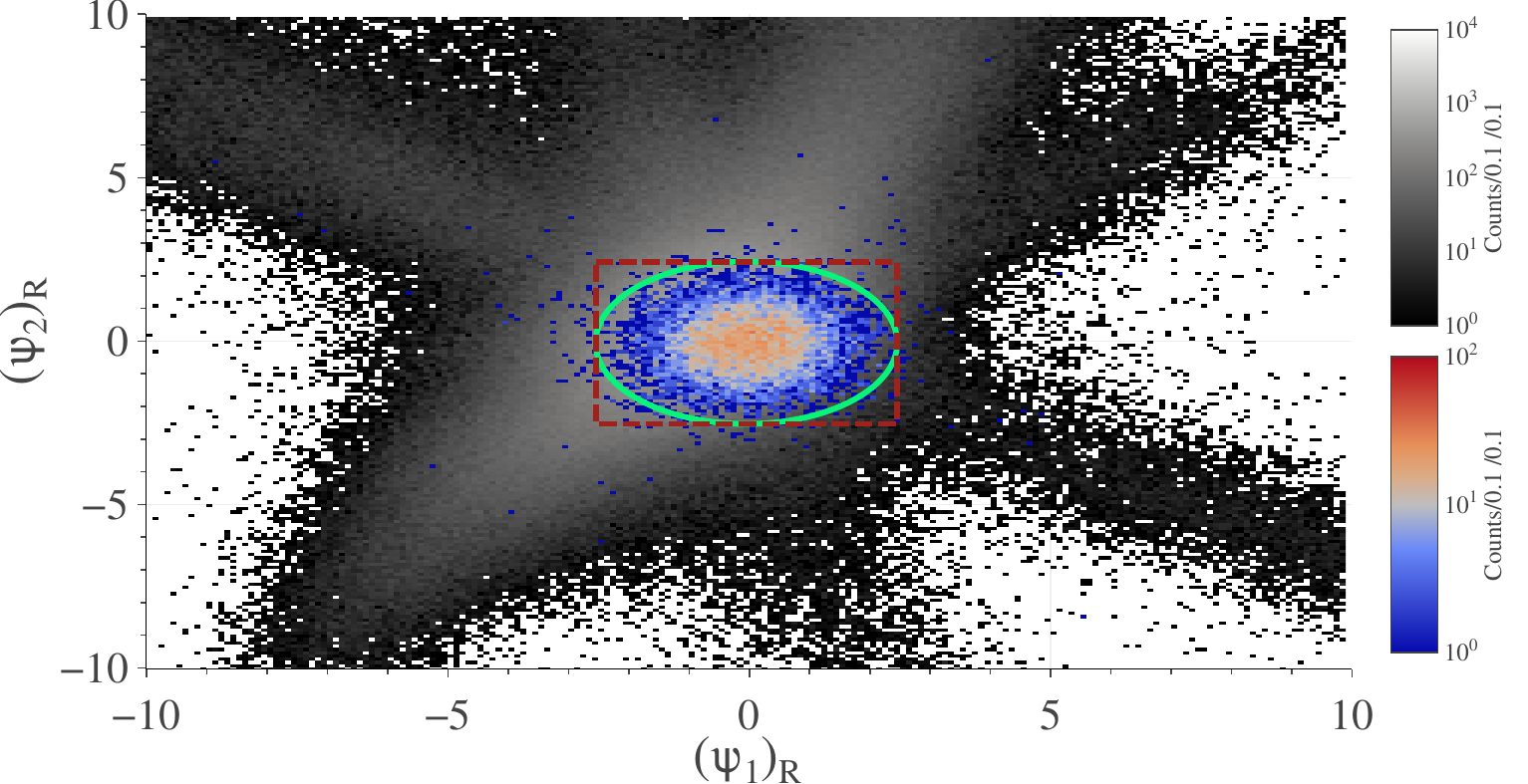}
\includegraphics[width=1.0\linewidth]{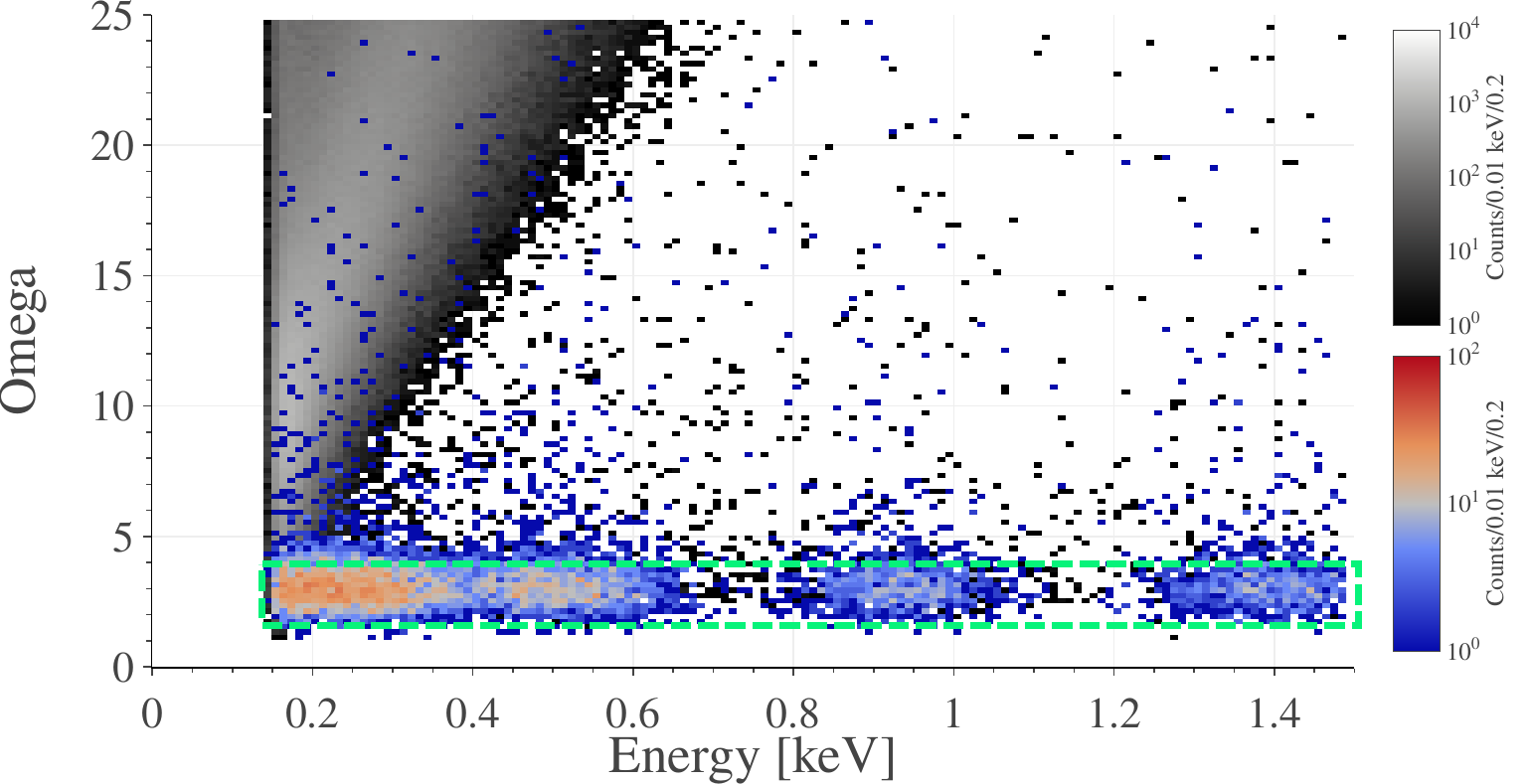}
\caption{{\bf Top:} 2-dimensional parameter space defined by the variables $(\psi_1)_R$ and $(\psi_2)_R$ for KID-47. The red-blue colorscale represents
LED events known to illuminate the correct die, while the gray points are unwanted background to be rejected.
The red rectangle represents the event selection achieved by independently requiring $|(\psi_1)_R| < 2.5$ and $|(\psi_2)_R| < 2.5$, which yields
high efficiency at the cost of moderate rejection power. The event selection achievable by defining the variable \mycolor{$\Omega^* = {\sqrt{ \sum_{j=1}^{2} \left( M_{i j} \psi_{j} \right)^{2} }}$} and imposing the condition $\Omega^* < 2.5$,
represented as a green circle, provides a higher rejection of the background while retaining a comparable efficiency on the
LED events.
{\bf Bottom:} Distribution of the $\Omega$ variable computed for KID-47 for LED events (red-blue colorscale) and for background events (gray colorscale). The variable allows for good separation down to approximately 200~eV. The green band represents the event selection corresponding to $1.5<\Omega<4.0$, retaining most of the known good LED events while rejecting events resulting from phonon leakage generated in other dice.} 
\label{fig7}
\end{figure}
To reject events that produce a signal in the main die but are originating elsewhere we analyze the signal amplitude in the \mycolor{neighboring} KIDs, evaluated in coincidence with the maximum in the central KID. For the \mycolortris{energy} spectrum discussed in Sec.~\ref{sec:analysis} and in Ref.~\cite{bullkid2024} we 
define for each neighboring die the following variable
\begin{equation}
\label{psi_n}
\psi_n = \dfrac{A_n -A \cdot r_n}
{\sqrt{\sigma_{0,n}^2 + r^2_n\cdot\sigma_0^2}}
\end{equation}
where with $A$ ($A_n$) we indicate the pulse amplitude in the main ($n^{th}$) KID, 
$r_n = \langle A_n / A\rangle$ is the average relative phonon leakage amplitude and with $\sigma_{0}$ ($\sigma_{0,n}$) we indicate the noise standard deviation.

To allow for the maximum sensitivity to the phonon leakage distribution, we compute for KID-47 and 49 the variables $\psi_{1...14}$ evaluated on all of the other KIDs.
We then compute the correlation matrix $C$ as
\begin{equation}
C_{ij} = \langle\psi_i \psi_j\rangle - \langle\psi_i\rangle\langle\psi_j\rangle
\end{equation}
where the indices $i$ and $j$ run over the 14 side KIDs. The correlation matrix can be diagonalized with the opportune 
change-of-basis matrix $M$ as
\begin{equation}
C_{\text{\tiny{DIAG.}}} = M^T C M
\end{equation}
with the eigenvalues $\left(C_{\text{\tiny{DIAG.}}}\right)_{ii}$ representing the variance of the decorrelated $\psi_i$ variables,
indicated with $\left(\psi_i\right)_R$. Once the $M$ matrix and the $\psi_i$ variables are computed, the decorrelated variables
are obtained as 
\begin{equation}
\left(\psi_i \right)_R = {\sum^{N_{neigh}}_{j=1} M_{ij}\psi _j}
\end{equation}
with $N_{neigh} = 14 $ in the current configuration.
To select only energy depositions interacting in the desired die we previously rejected events that do not satisfy the
condition $|\psi_i| < T$ with $i \in [1,N_{neigh}]$, selecting an $N_{neigh}$-dimensional \mycolor{hypercube} with an edge length 
of $2T$. A more efficient discrimination of signals from background can be achieved by defining a distance from the origin of the 
$N_{neigh}$-dimensional parameter space defined by $\left(\psi_i\right)_R$~\cite{magistrale_maiello} 
as 
\begin{equation}
\Omega = \sqrt{\sum^{N_{neigh}}_{i=1}\left[ \left(\psi_i\right)_R \right]^2  }
\end{equation}
which allows to retain only events contained in an hypersphere of radius $\Omega^*$, as illustrated in Fig.~\ref{fig7} (Top).

The distribution of $\Omega$ for KID-47 is shown in Fig.~\ref{fig7} (Bottom) for both background and LED events: given the good
separation between the two datasets we chose to retain only events that satisfy the condition $1.5 < \Omega < 4.0$.
Both the pulse shape parameters and the cluster cut on $\Omega$ were optimized to maximize the rejection of the events acquired by a trigger on the negative side of the data stream, representative of the rate of noise false positives.
The same cuts, combined with the trigger acceptance, yield overall LED event efficiencies of ($47\pm 1$)\% and ($55\pm 1$)\% for the KID-47 and KID-49 clusters, respectively

\end{appendix}

\end{document}

%% file: bullkid_plugin_authors.tex
\newcommand{\AffI}{Dipartimento di Fisica, Sapienza Università di Roma, P. le A. Moro 2, 00185 Roma, Italy}
\newcommand{\AffII}{INFN Sezione di Roma, P.le A. Moro 2, 00185 Roma, Italy}
\newcommand{\AffIII}{Dipartimento di Fisica "Enrico Fermi", Università di Pisa, Largo Bruno Pontecorvo 3, 56127 Pisa, Italy}
\newcommand{\AffIV}{INFN Sezione di Pisa, Largo Bruno Pontecorvo 3, 56127 Pisa, Italy}
\newcommand{\AffV}{Instituto de Física, Universidad Nacional Autónoma de México, A.P. 20-364, Ciudad de México 01000, México}
\newcommand{\AffVI}{INFN Sezione di Ferrara, Via Saragat 1, 44122 Ferrara, Italy}
\newcommand{\AffVII}{Université Grenoble Alpes, CNRS, Grenoble INP, Institut Néel, 38000 Grenoble, France}
\newcommand{\AffVIII}{INFN Laboratori Nazionali del Gran Sasso, via G. Acitelli n. 22, 67100 Assergi (AQ) - Italy}
\newcommand{\AffIX}{Institute for Data Processing and Electronics, Karlsruhe Institute of Technology, Hermann-von-Helmholtz-Platz 1 76344, Eggenstein-Leopoldshafen - Germany}
\newcommand{\AffX}{INFN - TIFPA, Via Sommarive 14, 38123 Povo (Trento) Italy}
\newcommand{\AffXI}{Dipartimento di Fisica e Scienze della Terra, Università di Ferrara, Via Saragat 1, 44100 Ferrara - Italy}
\newcommand{\AffXIII}{Gran Sasso Science Institute, Viale F. Crispi, 7 67100 L'Aquila}
\newcommand{\AffXIV}{Now at: Università di Roma "Tor Vergata", Dipartimento di Fisica, Via della Ricerca Scientifica 1, 00133 Roma, Italy}

\author{D.~Delicato\orcidlink{0009-0005-0516-6872}}\email{daniele.delicato@neel.cnrs.fr}\affiliation{\AffI}\affiliation{\AffII}\affiliation{\AffVII}
\author{A.~Acevedo-Rentería\orcidlink{0009-0003-1055-9377}}\affiliation{\AffV}
\author{M.~Folcarelli\orcidlink{0009-0009-7799-2515}}\affiliation{\AffI}\affiliation{\AffII}
\author{G.~Del~Castello\orcidlink{0000-0001-7182-358X}}\affiliation{\AffII}
\author{M.~Cappelli\orcidlink{0009-0002-6148-5964}}\affiliation{\AffI}\affiliation{\AffII}
\author{L.~E.~Ardila-Perez\orcidlink{0000-0002-7485-8267}}\affiliation{\AffIX}
\author{L.~Bandiera\orcidlink{0000-0002-5537-9674}}\affiliation{\AffVI}
\author{C.~Bonomo\orcidlink{0009-0001-0336-3857}}\affiliation{\AffI}\affiliation{\AffII}
\author{M.~Calvo\orcidlink{0000-0002-8752-6325}}\affiliation{\AffVII}
\author{R.~Caravita\orcidlink{0000-0002-8189-8814}}\affiliation{\AffX}
\author{F.~Carillo\orcidlink{0000-0002-7563-8960}}\affiliation{\AffIV}
\author{F.~Cescato~}\affiliation{\AffVI}\affiliation{\AffXI}
\author{U.~Chowdhury\orcidlink{0000-0002-1857-2598}}\affiliation{\AffVII}
\author{D.~A.~Crovo\orcidlink{0009-0000-2521-8901}}\affiliation{\AffIX}
\author{A.~Cruciani\orcidlink{0000-0003-2247-8067}}\affiliation{\AffII}
\author{A.~D’Addabbo\orcidlink{0000-0003-2668-962X}}\affiliation{\AffVIII}
\author{M.~De~Lucia\orcidlink{0000-0002-0519-9149}}\affiliation{\AffIII}\affiliation{\AffIV}
\author{M.~del~Gallo~Roccagiovine\orcidlink{0009-0006-5861-7443}}\affiliation{\AffI}\affiliation{\AffII}
\author{F.~Ferraro\orcidlink{0000-0002-2031-7879}}\affiliation{\AffVIII}
\author{S.~Fu\orcidlink{0000-0001-8509-6424}}\affiliation{\AffVIII}
\author{R.~Gartmann\orcidlink{0000-0003-3791-0051}}\affiliation{\AffIX}
\author{M.~Giammei\orcidlink{0009-0006-9104-2055}}\altaffiliation{\AffXIV}\affiliation{\AffI}\affiliation{\AffII}
\author{M.~Grassi\orcidlink{0000-0002-5551-1145}}\affiliation{\AffIV}
\author{V.~Guidi\orcidlink{0000-0003-4001-8064}}\affiliation{\AffVI}\affiliation{\AffXI}
\author{D.~L.~Helis\orcidlink{0000-0002-9111-0541}}\affiliation{\AffVIII}
\author{T.~Lari\orcidlink{0009-0003-7836-4779}}\affiliation{\AffIII}\affiliation{\AffIV}
\author{L.~Malagutti\orcidlink{0000-0002-0996-3095}}\affiliation{\AffVI}
\author{A.~Mazzolari\orcidlink{0000-0003-0804-6778}}\affiliation{\AffVI}\affiliation{\AffXI}
\author{A.~Monfardini\orcidlink{0000-0001-5337-5533}}\affiliation{\AffVII}
\author{T.~Muscheid\orcidlink{0000-0002-1108-7784}}\affiliation{\AffIX}
\author{D.~Nicolò\orcidlink{0000-0001-6414-946X}}\affiliation{\AffIII}\affiliation{\AffIV}
\author{F.~Paolucci\orcidlink{0000-0001-8354-4975}}\affiliation{\AffIII}\affiliation{\AffIV}
\author{D.~Pasciuto\orcidlink{0000-0001-8088-9716}}\affiliation{\AffII}
\author{L.~Pesce\orcidlink{0009-0001-5659-4691}}\affiliation{\AffI}\affiliation{\AffII}
\author{C.~Puglia\orcidlink{0000-0003-1493-0148}}\affiliation{\AffIV}
\author{D.~Quaranta\orcidlink{0009-0000-2954-4456}}\affiliation{\AffI}\affiliation{\AffII}
\author{C.~M.~A.~Roda\orcidlink{0000-0002-3020-4114}}\affiliation{\AffIII}\affiliation{\AffIV}
\author{S.~Roddaro\orcidlink{0000-0002-4707-1434}}\affiliation{\AffIII}\affiliation{\AffIV}
\author{M.~Romagnoni\orcidlink{0000-0002-2775-6903}}\affiliation{\AffVI}
\author{G.~Signorelli\orcidlink{0000-0001-8262-8245}}\affiliation{\AffIII}\affiliation{\AffIV}
\author{F.~Simon\orcidlink{0000-0002-5978-0289}}\affiliation{\AffIX}
\author{A.~Tartari\orcidlink{0000-0003-3082-138X}}\affiliation{\AffIV}
\author{E.~Vázquez-Jáuregui\orcidlink{0000-0003-4195-0961}}\affiliation{\AffV}
\author{M.~Vignati\orcidlink{0000-0002-8945-1128}}\affiliation{\AffI}\affiliation{\AffII}
\author{K.~Zhao\orcidlink{0009-0001-3423-5882}}\affiliation{\AffVIII}\affiliation{\AffXIII}